\documentclass[12pt,twoside,fleqn]{article}

\usepackage{a4wide,cite}
\usepackage{amsmath,amssymb}
\usepackage{epsfig,rotating}
\usepackage{axodraw}

\numberwithin{equation}{section}
\allowdisplaybreaks[4]

\newcommand{\MS}{\ensuremath{\overline{\text{MS}}}}

\begin{document}

\begin{titlepage}

\vspace*{1cm}

\centerline{\Large\bf\boldmath Fermionic corrections to the interference}
\vspace{3.3mm}
\centerline{\Large\bf\boldmath of the electro- and chromomagnetic dipole}
\vspace{2mm}
\centerline{\Large\bf\boldmath operators in $\bar B\to X_s\gamma$ at $O(\alpha_s^2)$}
\vskip 2.5cm

\begin{center}
  {\bf Thorsten Ewerth}\\[2mm]
  {\sl Dip.\ Fisica Teorica, Univ.\ di Torino \& INFN Torino, I-10125 Torino, Italy}
\end{center}
\medskip

\vskip 2cm

\begin{abstract}
\noindent We calculate the virtual and bremsstrahlung fermionic corrections due to the interference of the
electro- and chromomagnetic dipole operators in the inclusive $\bar B\to X_s\gamma$ decay at $O(\alpha_s^2)$ and
present analytical results for both the total decay rate and the photon energy spectrum.
\end{abstract}

\end{titlepage}


\section{Introduction}

The present experimental world average of the branching ratio of $\bar{B}\to X_s\gamma$, which includes
measurements by CLEO, BaBar and Belle \cite{Chen:2001fja,Aubert:2007my,Abe:2008sx}, is performed by the
Heavy Flavor Averaging Group \cite{Barberio:2007cr} and, for photon energies $E_\gamma > 1.6\,{\rm GeV}$, is
given by
\begin{equation}\label{hfag}
  \text{Br}(\bar B\to X_s\gamma)=\left(3.52\pm 0.23\pm 0.09\right)\times 10^{-4}\,,
\end{equation}
where the errors are combined statistical and systematic and due to the extrapolation to the common lower-cut
in the photon energy, respectively. Moreover, the total uncertainty, being already below 7\%, is expected to
reduce down to 5\% at the end of the B-factory era.

In order to keep pace with the improving experimental accuracy the theoretical prediction of the
$\bar{B}\to X_s\gamma$ branching ratio has to be known at the next-to-next-to-leading order (NNLO) level.
A first estimate of the $\bar{B}\to X_s\gamma$ branching ratio at this level of accuracy has been presented in
\cite{Misiak:2006zs}. For $E_\gamma > 1.6\,$GeV it reads
\begin{equation}\label{estimate}
  \text{Br}(\bar B\to X_s\gamma) = (3.15\pm 0.23)\times 10^{-4}\,.
\end{equation}
This estimate includes the three-loop dipole operator matching conditions \cite{Misiak:2004ew}, the three-loop
mixing of the four-quark operators \cite{Gorbahn:2004my}, the three-loop mixing of the dipole operators
\cite{Gorbahn:2005sa}, and the four-loop mixing of the four-quark operators into the dipole operators
\cite{Czakon:2006ss}. Also the two-loop matrix elements of the electromagnetic dipole operator together
with the corresponding bremsstrahlung terms (at $m_c=0$)
\cite{Blokland:2005uk,Melnikov:2005bx,Asatrian:2006ph,Asatrian:2006sm}, as well as the three-loop matrix elements
of the four-quark operators within the so-called large-$\beta_0$ approximation \cite{Bieri:2003ue} have been taken
into account. Finally, in order to obtain estimates of these matrix elements (together with other still unknown
ones) at the physical value of the charm quark mass $m_c$ an interpolation in the latter has been performed
in \cite{Misiak:2006ab}.

We should mention here that there are several perturbative and non-perturbative effects that have not been
considered when deriving the estimate given in (\ref{estimate}). Some of them are already available in the
literature: the four-loop mixing of $O_1,\dots,O_6$ into $O_8$ \cite{Czakon:2006ss}; the bremsstrahlung
contributions of the $(O_2,O_2)$, $(O_2,O_7)$ and $(O_7,O_8)$-interference at $O(\alpha_s^2\beta_0)$
\cite{Ligeti:1999ea}; the exact charm quark mass dependence of the $(O_7,O_7)$-interference at $O(\alpha_s^2)$
\cite{Asatrian:2006rq}; the three-loop virtual corrections due to charm and bottom quark loop insertions into gluon
propagators in the $(O_1,O_7)$ and $(O_2,O_7)$-interference \cite{Boughezal:2007ny}; the updated knowledge of the
semileptonic normalization factor \cite{Gambino:2008fj,Melnikov:2008qs,Pak:2008qt}; photon energy cut-off related
effects \cite{Neubert:2004dd,Becher:2005pd,Becher:2006qw,Andersen:2006hr}; and estimates for the
$O(\alpha_s\Lambda_\text{QCD}/m_b)$ corrections \cite{Lee:2006wn}. Other effects are unknown at the moment, like the
complete virtual and bremsstrahlung contributions to the $(O_7,O_8)$ and $(O_8,O_8)$-interference at
$O(\alpha_s^2)$ (only the contribution of the $(O_7,O_8)$-interference at $O(\alpha_s^2 \beta_0)$ is known
\cite{Bieri:2003ue,Ligeti:1999ea}), and of course the exact $m_c$-dependence of various matrix elements beyond
the large-$\beta_0$ approximation, in order to improve (or even remove) the uncertainty due to the interpolation
in $m_c$ \cite{Misiak:2006ab}. The individual contributions listed above are all expected to remain within the
uncertainty given in (\ref{estimate}), nevertheless they should be taken into account in  future updates.

In the present paper we repeat the calculation of the $(O_7,O_8)$-interference contribution performed in
\cite{Bieri:2003ue,Ligeti:1999ea} and extend it to include not only the effects of massless quark loops but
also those due to massive ones. More precisely, we calculate those $O(\alpha_s^2)$ contributions which can be
obtained from the Feynman diagrams contributing to the $(O_7,O_8)$-interference at $O(\alpha_s)$ when dressing
the gluon propagators with massless up, down and strange quark loops as well as with massive charm and bottom
quark loops. We work out the effects of these contributions to the photon energy spectrum
$d\Gamma(b\to X^{\rm partonic}_s\gamma)/dE_\gamma$ and to the total decay width
$\Gamma(b\to X^{\rm partonic}_s\gamma)|_{E_\gamma>E_0}$, where $E_0$ denotes the cut in the photon energy.

The organization of this paper is as follows. In section 2 we present our final results for the total decay
width and the photon energy spectrum and describe briefly the calculation of the relevant Feynman diagrams.
The numerical impact of the fermionic corrections on $\text{Br}(\bar B\to X_s\gamma)$ is estimated in section
3. Finally, we summarize in section 4.


\section{Fermionic corrections}\label{sec_2}

Within the low-energy effective theory the partonic $b\to X_s\gamma$ decay rate can be written as
\begin{equation}\label{decay_rate}
\Gamma(b\to X_s^{\rm parton}\gamma)_{E_\gamma>E_0} = \frac{G_F^2\alpha_{\rm em}
 \overline{m}_b^2(\mu)m_b^3}{32\pi^4}\,|V_{tb}^{}V_{ts}^*|^2\,\sum_{i,j}C_i^{\rm eff}(\mu)\,
 C_j^{\rm eff}(\mu)\,G_{ij}(E_0,\mu)\,,
\end{equation}
where $m_b$ and $\overline{m}_b(\mu)$ denote the pole and the running $\MS$ mass of the $b$ quark, respectively,
$C_i^{\rm eff}(\mu)$ the effective Wilson coefficients at the low-energy scale, and $E_0$ the energy cut in the
photon spectrum.\footnote{In this paper we assume that the products $C_i^{\rm eff}(\mu)\,C_j^{\rm eff}(\mu)$ are
real. That is our formulas are not applicable to physics scenarios beyond the standard model which produce
complex short distance couplings.}

As already anticipated in the introduction, we will focus on the function $G_{78}(E_0,\mu)$ corresponding to the
interference of the electro- and chromomagnetic dipole operators
\begin{align}
 O_7 &= \frac{e}{16\pi^2}\,\overline{m}_b(\mu)\left(\bar s\sigma^{\mu\nu}P_Rb\right)F_{\mu\nu}
\end{align}
and
\begin{align}
 O_8 &= \frac{g}{16\pi^2}\,\overline{m}_b(\mu)\left(\bar s\sigma^{\mu\nu}P_RT^ab\right)G_{\mu\nu}^a\,,
\end{align}
respectively. In NNLO approximation this function can be decomposed as follows,
 \begin{align}\label{g77_exp}
  G_{78}(E_0,\mu) &= \frac{\alpha_s(\mu)}{4\pi}\,C_F Y^{(1)}(z_0,\mu) +
  \left(\frac{\alpha_s(\mu)}{4\pi}\right)^2 C_F Y^{(2)}(z_0,\mu) + O(\alpha_s^3)\,,
\end{align}
where
\begin{align}\label{colorf}
  Y^{(2)}(z_0,\mu) &= C_F Y^{(2,\mbox{{\tiny CF}})}(z_0,\mu) +
  C_A Y^{(2,\mbox{{\tiny CA}})}(z_0,\mu)\nonumber\\[2mm]
  &\qquad+ T_R N_L Y^{(2,\mbox{{\tiny NL}})}(z_0,\mu) + T_R N_H Y^{(2,\mbox{{\tiny NH}})}(z_0,\mu) +
    T_R N_V Y^{(2,\mbox{{\tiny NV}})}(z_0,\mu)\,.
\end{align}
Here, $z_0=2E_0/m_b$, $N_L$, $N_H$ and $N_V$ denote the number of light ($m_q=0$), heavy ($m_q=m_b$),
and purely virtual ($m_q=m_c$) quark flavors, respectively, $\alpha_s(\mu)$ is the running coupling constant
in the $\MS$ scheme, and $C_F$, $C_A$ and $T_R$ are the color factors with numerical values given by 4/3, 3 and
1/2, respectively. In this paper we present results for the functions $Y^{(2,i)}(z_0,\mu)$ with
$i=\mbox{\small NL}$, $\mbox{\small NH}$, $\mbox{\small NV}$.  The calculation of the functions
$Y^{(2,i)}(z_0,\mu)$ with $i=\mbox{\small CF}$, $\mbox{\small CA}$ is the subject of another publication
\cite{workinprogress}.

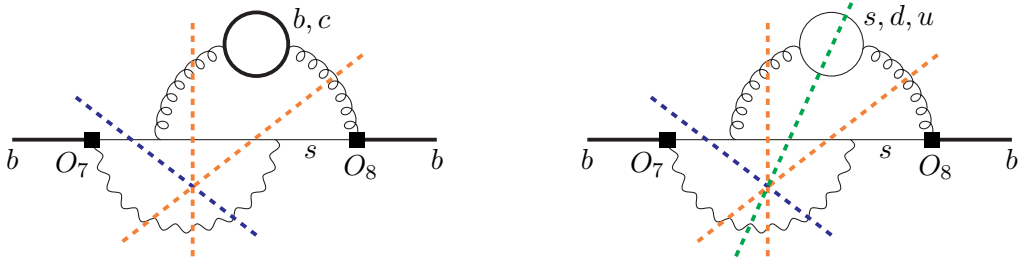
\begin{figure}[t]
  \vspace*{2cm}
  \begin{center}
    \begin{tabular}{ccc}
      \begin{picture}(0,0)(0,0)
        \SetScale{.8}
        \SetWidth{1.8}
        \Line(-100,0)(-60,0)
        \Line(100,0)(60,0)
        \CArc(15,45)(15,0,360)
        \SetWidth{.5}
        \Line(-60,0)(60,0)
        \PhotonArc(-18,0)(42,-180,0){2}{12.5}
        \GlueArc(15,0)(45,0,71){3}{7}
        \GlueArc(15,0)(45,109,180){3}{7}
        \Text(40,50)[tr]{\small $b,c$}
        \Text(-78,-11)[br]{\small $b$}
        \Text(78,-11)[bl]{\small $b$}
        \Text(30,-7)[bl]{\small $s$}
        \Text(-50,0)[c]{\rule{2mm}{2mm}}
        \Text(-51,-14)[br]{\small $O_7$}
        \Text(50,0)[c]{\rule{2mm}{2mm}}
        \Text(45,-15)[bl]{\small $O_8$}
        \SetWidth{1.8}
        \SetColor{Blue}
        \DashLine(-70,20)(15,-45){3}
        \SetColor{Orange}
        \DashLine(-15,55)(-15,-55){3}
        \DashLine(-48,-48)(65,40){3}
      \end{picture}
      & \hspace{6.8cm} &
      \begin{picture}(0,0)(0,0)
        \SetScale{.8}
        \SetWidth{1.8}
        \Line(-100,0)(-60,0)
        \Line(100,0)(60,0)
        \SetWidth{.5}
        \CArc(15,45)(15,0,360)
        \Line(-60,0)(60,0)
        \PhotonArc(-18,0)(42,-180,0){2}{12.5}
        \GlueArc(15,0)(45,0,71){3}{7}
        \GlueArc(15,0)(45,109,180){3}{7}
        \Text(51,51)[tr]{\small $s,d,u$}
        \Text(-78,-11)[br]{\small $b$}
        \Text(78,-11)[bl]{\small $b$}
        \Text(30,-7)[bl]{\small $s$}
        \Text(-50,0)[c]{\rule{2mm}{2mm}}
        \Text(-51,-14)[br]{\small $O_7$}
        \Text(50,0)[c]{\rule{2mm}{2mm}}
        \Text(45,-15)[bl]{\small $O_8$}
        \SetWidth{1.8}
        \SetColor{Blue}
        \DashLine(-70,20)(15,-45){3}
        \SetColor{Orange}
        \DashLine(-15,55)(-15,-55){3}
        \DashLine(-48,-48)(65,40){3}
        \SetColor{Green}
        \DashLine(-30,-55)(27,70){3}
      \end{picture}\\[12mm]
    \end{tabular}
  \end{center}
  \caption{\sl Two sample $b$ quark selfenergy diagrams which are proportional to the number of light, heavy
    and purely virtual quark flavors and whose 2-, 3- and 4-particle cuts contribute to the $b\to s\gamma$ (blue
    dashed lines), $b\to s\gamma g$ (orange dashed lines), and $b\to s\gamma q\bar q$ (green dashed line)
    transitions at $O(\alpha_s^2)$. See text for more details.}\label{sample_diags}
\end{figure}

The functions $Y^{(2,i)}$ with $i=\mbox{\small NL}$, $\mbox{\small NH}$, $\mbox{\small NV}$ appearing in
(\ref{colorf}) receive contributions from the $b\to s\gamma$, $b\to s\gamma g$, and $b\to s\gamma q\bar q$
($q\in\{u,d,s\}$, $m_q=0$) transitions. The latter are contained in the $b$ quark selfenergies which arise
from those at $O(\alpha_s)$ when dressing the gluon propagators with massless and massive quark
loops.\footnote{There are also $b$ quark selfenergies where the photon runs from $O_7$ to the quark loop
and which cannot be obtained from those at $O(\alpha_s)$. However, it is easy to show that these diagrams
cancel amongst themselves without performing loop and phase-space integrations (see Furry's theorem).} Two sample
$b$ quark self-energies containing cuts with two, three and four particles in the intermediate state are displayed
in figure \ref{sample_diags}. As far as the diagrams containing massive quarks in the fermion loop are concerned,
like, e.g., the one given on the left-hand side of figure \ref{sample_diags}, we do not have to calculate cuts
with four particles in the intermediate state since such cuts would run through the bottom or charm quark loop
and (i) $b$ quarks are of course kinematically not allowed to appear in the final state and (ii) events involving
charmed hadrons in the final state are not included on the experimental side. On the other hand, for the diagrams
containing massless quarks in the fermion loop like, e.g., the one given on the right-hand side of figure
\ref{sample_diags}, the contributions from $q\bar q$ production ($q\in\{u,d,s\}$), that is four-particle cuts
running through massless quark loops, have to be taken into account.

We work in $d=4-2\epsilon$ space-time dimensions to regularize ultraviolet, infrared and collinear
singularities, and adopt the renormalization prescription from \cite{Asatrian:2006ph,Asatrian:2006rq}.
Most of the renormalization constants necessary to render our results ultraviolet finite can be found there.
The only exceptions are those which describe the selfmixing of $O_8$ at one-loop and the mixing of $O_8$ into
$O_7$ up to two-loops; they can be extracted from \cite{Misiak:1994zw}. For some technical details concerning the
evaluation of the two-loop integrals involving both the bottom and charm quark mass $m_b$ and $m_c$,
respectively, we refer the reader to the end of this section.

In order to obtain a compact presentation of our findings we split the functions $Y^{(2,i)}$ with
$i=\mbox{\small NL}$, $\mbox{\small NH}$, $\mbox{\small NV}$ into two parts, namely
\begin{align}\label{Ysplit}
 Y^{(2,i)}(z_0,\mu) &= Y^{(2,i)}(0,\mu) - \delta Y^{(2,i)}(z_0,\mu),
\end{align}
where the first terms give always the contribution to the full inclusive decay rate, and the second ones
correct for the fact that in the experiments a lower cut in the photon energy is applied. Performing the same
splitting for the function $Y^{(1)}$ appearing in (\ref{g77_exp}), our findings for the two individual
contributions at $O(\alpha_s)$ read
\begin{align}
 Y^{(1)}(0,\mu) &= \frac{4}{9}\left(29-2\pi^2\right)+\frac{16}{3}\,L_\mu\,,\\[3mm]
 \delta Y^{(1)}(z_0,\mu) &= \frac{2}{9}\,z_0\left(z_0^2+24\right)-\frac{8}{3}(z_0-1)\ln(1-z_0) -
   \frac{8}{3}\,\text{Li}_2(z_0)\,,\label{Ynlo}
\end{align}
while those at $O(\alpha_s^2)$ are given by
\begin{align}
 Y^{(2,\mbox{{\tiny NL}})}(0,\mu) &= -\frac{16}{81}\left(328-13\pi^2\right) -
   \frac{64}{27}\left(18-\pi^2\right)L_\mu-\frac{64}{9}\,L_\mu^2+\frac{64}{3}\,\zeta_3\,,\label{Y0nl}\\[3mm]
 Y^{(2,\mbox{{\tiny NH}})}(0,\mu) &= \frac{8}{81}\left(244-27\sqrt{3}\,\pi-61\pi^2\right) -
   \frac{64}{27}\left(18-\pi^2\right)L_\mu\nonumber\\[2mm]
 &\qquad -\frac{64}{9}\,L_\mu^2-\frac{64}{27}\,\zeta_3 +
   32\sqrt{3}\,\text{Cl}_2\left(\frac{\pi}{3}\right)\,,\label{Y0nh}\\[3mm]
 Y^{(2,\mbox{{\tiny NV}})}(0,\mu) &=
   -\frac{16}{81}\left[157-279\rho-\pi^2\left(5+9\rho^2-42\rho^{3/2}\right)\right] -
   \frac{64}{27}\left(18-\pi^2\right)L_\mu\nonumber\\[2mm]
 &\qquad -\frac{64}{9}\,L_\mu^2+\frac{16}{27}\left(22-\pi^2+10\rho\right)\ln\rho +
   \frac{16}{27}\left(8+9\rho^2\right)\ln^2\!\rho\nonumber\\[2mm]
 &\qquad -\frac{16}{27}\,\ln^3\!\rho-\frac{8}{9}\left(1-6\rho^2\right)\Phi_1(\rho) -
   \frac{8}{27}(19-46\rho)\,\Phi_2(\rho)\nonumber\\[2mm]
 &\qquad -\frac{32}{27}(13+14\rho)\,\Phi_3(\rho)-\frac{64}{9}\,\Phi_4(\rho) -
   \frac{32}{9}\,\ln\rho\,\text{Li}_2(1-\rho)\nonumber\\[2mm]
 &\qquad +\frac{32}{27}\left(5+9\rho^2+14\rho^{3/2}\right)\text{Li}_2(1-\rho) -
   \frac{1792}{27}\rho^{3/2}\,\text{Li}_2\left(1-\sqrt{\rho}\right)\nonumber\\[2mm]
 &\qquad +\frac{64}{9}\,\text{Li}_3(1-\rho)+\frac{64}{9}\,
   \text{Li}_3\left(1-\frac{1}{\rho}\right)\,,\label{Y0nv}\\[5mm]
 \delta Y^{(2,\mbox{{\tiny NL}})}(z_0,\mu) &= -\frac{4}{27}\,z_0\left(7 z_0^2-17 z_0+238\right) -
   \frac{8}{3}\,\delta Y^{(1)}(z_0,\mu)L_\mu\nonumber\\[2mm]
 &\qquad +\frac{8}{27}\left(z_0^3-6 z_0^2+80 z_0-75+6\pi ^2\right)\ln(1-z_0)\nonumber\\[2mm]
 &\qquad -\frac{16}{3}(z_0-1)\ln^2(1-z_0)-\frac{16}{3}\,\ln z_0\ln^2(1-z_0)\nonumber\\[2mm]
 &\qquad -\frac{32}{27}(3 z_0-8)\,\text{Li}_2(z_0)-\frac{32}{3}\,\ln(1-z_0)\,\text{Li}_2(z_0)\nonumber\\[2mm]
 &\qquad +\frac{32}{9}\,\text{Li}_3(z_0)-\frac{32}{3}\,\text{Li}_3(1-z_0) +
   \frac{32}{3}\,\zeta_3\,,\label{Yz0nl}\\[3mm]
 \delta Y^{(2,\mbox{{\tiny NH}})}(z_0,\mu) &= -\frac{8}{3}\,\delta Y^{(1)}(z_0,\mu)L_\mu\,,\label{Yz0nh}\\[3mm]
 \delta Y^{(2,\mbox{{\tiny NV}})}(z_0,\mu) &= -\frac{4}{3}\,\delta Y^{(1)}(z_0,\mu)(2L_\mu-\ln\rho)\,.\label{Yz0nv}
\end{align}
In writing these equations we introduced the short-hand notations
\begin{equation}
  \rho=\frac{m_c^2}{m_b^2}\qquad\mbox{and}\qquad L_\mu=\ln\!\left(\frac{\mu}{m_b}\right)\,.
\end{equation}
The definitions of the auxiliary functions $\Phi_n(\rho)$ as well as those of the polylogarithms
$\mbox{Li}_n(z)$ and the Clausen function $\mbox{Cl}_2(z)$ can be found in appendix \ref{app_b}. The
numerical value of the Clausen function at $z=\pi/3$ is approximately given by $1.014942$, and
$\zeta_3\approx 1.202057$ is equal to Riemann's theta functions $\zeta(n)$ at $n=3$. Equations (\ref{Y0nv})
and (\ref{Yz0nv}) hold for $\rho>0$.

Turning now to our findings for the photon energy spectrum, we rewrite the function $G_{78}(E_0,\mu)$ as an
integral over the (rescaled) photon energy,
\begin{equation}
  G_{78}(E_0,\mu) = \int_{z_0}^1\!dz\,\frac{dG_{78}(z,\mu)}{dz}\,,\qquad z=\frac{2E_\gamma}{m_b}\,.
\end{equation}
In NNLO approximation the integrand can be written as follows,
\begin{align}
 \frac{dG_{78}(z,\mu)}{dz} &= \frac{\alpha_s(\mu)}{4\pi}\,C_F\widetilde Y^{(1)}(z,\mu) +
   \left(\frac{\alpha_s(\mu)}{4\pi}\right)^2C_F\widetilde Y^{(2)}(z,\mu) + O(\alpha_s^3)\,,
\end{align}
where, in analogy to (\ref{colorf}),
\begin{align}\label{colorf_spec}
 \widetilde Y^{(2)}(z,\mu) = T_RN_L\widetilde Y^{(2,\mbox{{\tiny NL}})}(z,\mu) +
   T_RN_H\widetilde Y^{(2,\mbox{{\tiny NH}})}(z,\mu) +
   T_RN_V\widetilde Y^{(2,\mbox{{\tiny NV}})}(z,\mu) + \dots\,,
\end{align}
with the ellipses denoting terms which are proportional to the colorfactors $C_F$ and $C_A$. The
next-to-leading order (NLO) function $\widetilde Y^{(1)}(z,\mu)$ is given by
\begin{align}\label{Ynlo_spec}
 \widetilde Y^{(1)}(z,\mu) = Y^{(1)}_{2\text{-cuts}}(0,\mu)\,\delta(1-z) +
   \frac{d}{dz}\,\delta Y^{(1)}(z,\mu)\,,
\end{align}
where $\delta Y^{(1)}(z,\mu)$ can be obtained from (\ref{Ynlo}) by replacing $z_0$ by $z$, and
$Y^{(1)}_{2\text{-cuts}}(0,\mu)$ can be found in appendix \ref{app_a}. The latter function summarizes the
contribution of all 2-particle cuts entering the functions $Y^{(1)}(z_0,\mu)$. The terms proportional to $N_L$,
$N_H$ and $N_V$ appearing in (\ref{colorf_spec}) can be written in complete analogy to (\ref{Ynlo_spec}),
\begin{align}\label{Yi_spec}
 \widetilde Y^{(2,i)}(z,\mu) = Y^{(2,i)}_{2\text{-cuts}}(0,\mu)\,\delta(1-z) +
   \frac{d}{dz}\,\delta Y^{(2,i)}(z,\mu)\,,
\end{align}
with $\delta Y^{(2,i)}(z,\mu)$ given in (\ref{Yz0nl})--(\ref{Yz0nv}) and $Y^{(2,i)}_{2\text{-cuts}}(0,\mu)$
in appendix \ref{app_a}. Since the contributions of the 2-particle cuts are by themselves free of infrared and
collinear singularities it was not necessary to introduce plus-distributions in (\ref{Yi_spec})
and (\ref{Ynlo_spec}).

We remark that the terms proportional to $N_L$, that is the functions $Y^{(2,\mbox{{\tiny NL}})}(0,\mu)$ and
$\delta Y^{(2,\mbox{{\tiny NL}})}(z_0,\mu)$, are already known in the literature \cite{Bieri:2003ue,Ligeti:1999ea}
and we completely agree with the results given there. The functions $Y^{(2,i)}(z_0,\mu)$ with
$i=\mbox{\small NH}$, $\mbox{\small NV}$ are however new.\footnote{I would like to thank Christoph Greub for
checking equation (\ref{Y2cutsnh}), which contains the contribution of the 2-particle cuts being proportional to
$N_H$, numerically.}

In the remainder of this section we will summarize the technical details of the calculation. However, we refrain
from repeating the algebraic reduction procedure of the 2-, 3- and 4-particle cuts of the three-loop $b$ quark
selfenergies to a set of so-called master integrals as well as from discussing appropriate parametrizations of
the phase-space integrals here since this has already been done in great detail in
\cite{Asatrian:2006ph,Asatrian:2006sm}. Instead, we will briefly describe how we solved the non-trivial two-loop
integrals involving the two mass scales $m_b$ and $m_c$. First, we introduced Feynman parameters in the standard
way and performed the loop-integrations. Subsequently, we applied the Mellin-Barnes technique
\cite{Smirnov:1999gc,Tausk:1999vh} based on the relation
\begin{equation}
  \frac{1}{(x+y)^\lambda} = \frac{1}{\Gamma(\lambda)}\int\limits_C\!\frac{ds}{2\pi i}
  \frac{x^s}{y^{\lambda+s}}\,\Gamma(-s)\Gamma(\lambda+s)\,,
\end{equation}
where the integration contour $C$ runs from $-i\infty$ to $+i\infty$ such that it separates the poles generated
by the two $\Gamma$ functions. In this way all powers of a sum of several terms could be replaced by one- or
two-fold Mellin-Barnes integrals which made the integration over the Feynman parameters trivial. Finally, we
closed the integration contours $C$ sidewards by a half-circle with infinite radius and summed up the
enclosed residues. In our case all infinite sums involving the mass ratio $m_c/m_b$ could be reduced to the
inverse binomial sums given in \cite{Davydychev:2003mv}, and we obtained solutions for all two-loop integrals
which are valid for arbitrary values of $m_c/m_b$. We checked our analytical results for the master integrals
for several values of $m_c/m_b$ by numerically integrating over the Feynman parameter representations.

Two other checks of our calculation are provided by taking the limits $m_c\to0$ and $m_c\to m_b$ of the
contributions of the 2-particle cuts proportional to $N_V$,
\begin{align}
  \lim_{\rho\to 0}Y^{(2,\mbox{{\tiny NV}})}_{2\text{-cuts}}(0,\mu) &=
    Y^{(2,\mbox{{\tiny NL}})}_{2\text{-cuts}}(0,\mu)\,,\nonumber\\[2mm]
  \lim_{\rho\to 1}Y^{(2,\mbox{{\tiny NV}})}_{2\text{-cuts}}(0,\mu) &=
    Y^{(2,\mbox{{\tiny NH}})}_{2\text{-cuts}}(0,\mu)\,,
\end{align}
which reproduce our results proportional to $N_L$ and $N_H$. Also the limit $\rho\to 1$ of the complete expression
$Y^{(2,i)}(z_0,\mu)$ for $i=\mbox{NV}$ reduces to that for $i=\mbox{NH}$. We note, however, that it is not possible
to take the limit $\rho\to 0$ of the complete expression for $i=\mbox{NV}$ since we excluded the contributions with
massive $c\bar c$-pairs in the final state, and hence some $\ln\rho$ terms being present in
$Y^{(2,\mbox{{\tiny NV}})}(z_0,\mu)$ parametrize infrared and collinear singularities. The last check concerns the
asymptotic behavior for $m_c\gg m_b$. In this limit our result for the complete expression with $i=\mbox{NV}$
reduces to
\begin{align}
  Y^{(2,\mbox{{\tiny NV}})}(z_0,\mu) = \frac{364}{81} -
    \left[\frac{224}{27}+\frac{8}{3}\,Y^{(1)}(z_0,\mu)\right]\ln\!\left(\frac{\mu}{m_c}\right) +
    \frac{64}{9}\,\ln^2\!\left(\frac{\mu}{m_c}\right)+O\left(\frac{1}{\rho}\right)\,,
\end{align}
which is in agreement with the asymptotic form found in \cite{Misiak:2004ew} (see equation (5.10) of that
reference).


\section{Numerical impact}

\begin{figure}[t]
  \begin{center}
  \begin{tabular}{@{\hspace{5mm}}c@{\hspace{10mm}}c}
    \epsfig{figure=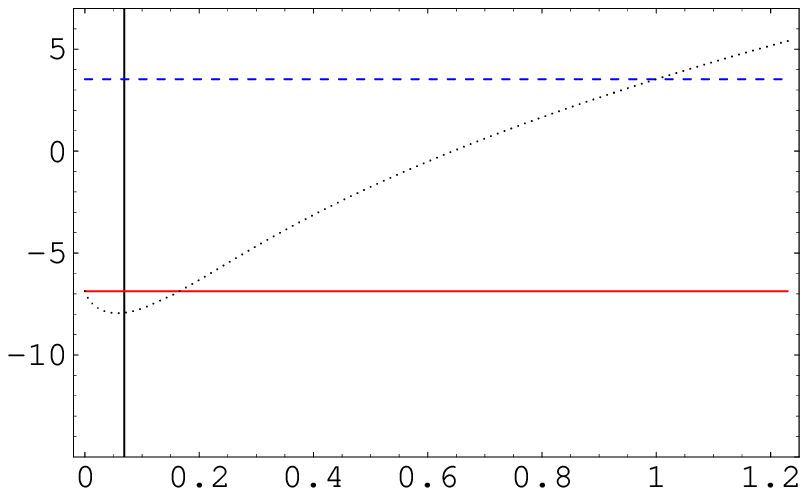,height=4.5cm} &
    \epsfig{figure=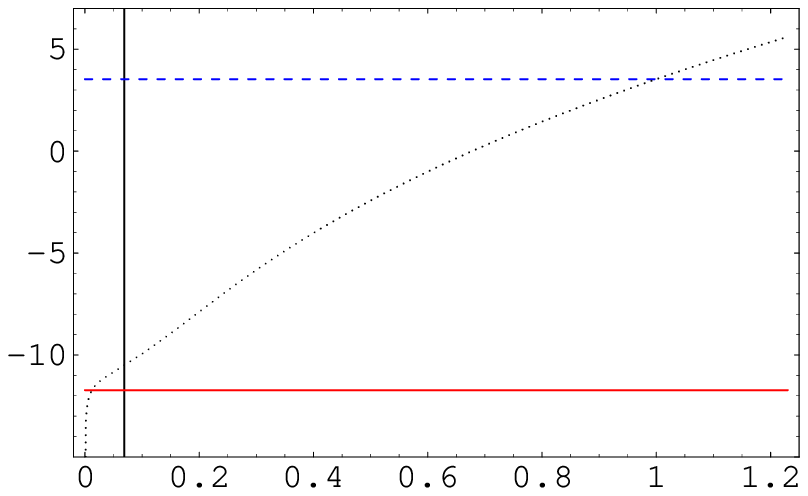,height=4.5cm}\\[-2mm]
    \hspace*{.5cm}{\small $\rho$} & \hspace*{.6cm}{\small $\rho$}\\
    \hspace*{-7.5cm}\begin{rotate}{90}\hspace*{23mm}{\small $Y^{(2,i)}_{2\text{-cuts}}(0,m_b)$}\end{rotate} &
    \hspace*{-7.5cm}\begin{rotate}{90}\hspace*{23mm}{\small $Y^{(2,i)}(z_0,m_b)$}\end{rotate}\\
  \end{tabular}\\[-5mm]
  \caption{\sl $Y^{(2,i)}_{2\text{-cuts}}(0,m_b)$ (left) and $Y^{(2,i)}(z_0,m_b)$ (right) as a function of
    $\rho=m_c^2/m_b^2$ for $i=\mbox{NL}$ (red solid line), $i=\mbox{NH}$ (blue dashed line) and $i=\mbox{NV}$
    (black dotted curve). The vertical lines indicates the physical value $\rho\approx(0.262)^2.$}\label{plot}
  \end{center}
\end{figure}

In the left frame of figure \ref{plot} we show the dependence of the functions $Y^{(2,i)}_{2\text{-cuts}}(0,m_b)$
on the mass ratio $\rho=m_c^2/m_b^2$ for $i=\mbox{NL}$, $\mbox{NH}$, $\mbox{NV}$. As can be seen, the function
for $i=\mbox{NH}$ (blue dashed line) differs from that for $i=\mbox{NL}$ (red solid line) by a factor of about
$-0.5$. On the other hand, at the physical value $\rho\approx(0.262)^2$, the function for $i=\mbox{NV}$
(black dotted curve) has a smaller value of about 15\% compared to that for $i=\mbox{NL}$.

The right frame of figure \ref{plot} displays the functions $Y^{(2,i)}(z_0,m_b)$ for $i=\mbox{NL}$, $\mbox{NH}$,
$\mbox{NV}$ as functions of the mass ratio $\rho$. They differ from the ones shown in the left frame by adding
the contributions from the 3- and 4-particle cuts, with the latter depending on the (rescaled) photon energy
cutoff $z_0$. The numerical value we chose in this illustration is given by $z_0=0.68$ and corresponds to
$E_0=1.6\,$GeV. As seen, the main effect of the bremsstrahlung corrections is to shift the function for
$i=\mbox{NL}$ (red solid line) and $i=\mbox{NV}$ (black dotted curve) down by a factor of about 1.7 and 1.3,
respectively. Also the aforementioned logarithmic singularity for $i=\mbox{NV}$ can be observed for $\rho\to 0$.
There is no shift for $i=\mbox{NH}$ (blue dashed line) since we set $\mu$ equal to $m_b$ in our illustration
(see (\ref{Yz0nh})).

We remark that other values for the renormalization scale $\mu$ than $m_b$ lead merely to a
shift of the three curves plotted in the left frame of figure \ref{plot} by the same amount up or down. The
reason for this is that the three quantities given in (\ref{Y2cutsnl})--(\ref{Y2cutsnv}) have exactly the same
$\mu$-dependence. The same comment is also true for the three curves shown in the right frame of figure \ref{plot},
as can be seen from (\ref{Y0nl})--(\ref{Yz0nv}). However, it is clear that a variation of $\mu$ will change the
relative importance with which each individual contribution enters the function $G_{78}(E_0,\mu)$.
For example, the choice $\mu=1.2\,m_b$ leads to $Y^{(2,\mbox{{\tiny NH}})}(z_0,\mu)\approx 0$, and hence the
fermionic corrections will be dominated by the two functions with $i=\mbox{NL}$ and $i=\mbox{NV}$. On the other
hand, for smaller values of $\mu$ than $m_b$ the situation can be reversed.

Next, we compare the fermionic corrections at NNLO with the NLO result. Using the numerical values
$\alpha_s(m_b)=0.22$, $N_L=3$, $N_H=1$ and $N_V=1$ (for the other input parameters we use the same values as
before), we find
\begin{equation}
  G_{78}(E_0,m_b) = 0.086 - 0.009 + \ldots = 0.077 + \ldots\,,
\end{equation}
where the two numbers after the first equality sign correspond to the $O(\alpha_s)$ and the fermionic
$O(\alpha_s^2)$ contributions given in (\ref{g77_exp}), and the ellipses denote the still unknown
$O(\alpha_s^2)$ terms proportional to $C_F$ and $C_A$ as well as higher order corrections. Thus, at $\mu=m_b$,
the effect of the NNLO fermionic corrections is to lower the NLO value of $G_{78}(E_0,m_b)$ by around 10\%.
For $\mu=2.5$ and $7.5\,$GeV, the $O(\alpha_s)$ term in $G_{78}(E_0,\mu)$ changes to 0.009 and 0.124,
respectively, and the $O(\alpha_s^2)$ correction shifts these values by around 3\% and -11\%, respectively.

Finally, we estimate the effect of the fermionic corrections at $O(\alpha_s^2)$ on the branching ratio of
$\bar B\to X_s\gamma$. As seen in figure \ref{plot}, the contributions with massive charm quark loops
($i=\mbox{NV}$) at the physical value $\rho\approx(0.262)^2$ are of comparable size as those with massless
quarks in the loops ($i=\mbox{NL}$). Hence, the charm quark mass effects can with good accuracy be described
by a single massless quark entering the large-$\beta_0$ approximation. On the other hand, approximating the
contributions due to massive bottom quark loops ($i=\mbox{NH}$) by massless ones is not very accurate (see
figure \ref{plot}). Here, however, one should bear in mind that in the physical application we have three
massless and only two massive quarks. That is the leading correction to $\text{Br}(\bar B\to X_s\gamma)$ will
be given by the sum of the contributions with $i=N_L$ and $i=N_V$, where the former is weighted by a factor
of three, and the correction due to the contribution with $i=N_H$ will only appear at the subleading level.
In fact, the exact result of the fermionic corrections at $\mu=m_b$ can be accurately approximated by setting
$N_L=3.6$ and $N_H=N_V=0$ in (\ref{colorf}). Thus, the effect of the massive bottom quark loops at $\mu=m_b$ can
be accounted for in the large-$\beta_0$ approximation by reducing the number of massless quark flavors by -0.4.
Given that the large-$\beta_0$ corrections of the $(O_7,O_8)$-interference affect the branching ratio of
$\bar B\to X_s\gamma$ by around 0.7\% for $\mu=m_b$, we conclude that this will not be altered drastically when
implementing the exact results for the fermionic corrections with massive quarks.\footnote{Here we should
mention that the effect of the large-$\beta_0$ corrections in the $(O_7,O_8)$-interference on the branching
ratio of $\bar B\to X_s\gamma$ stays below 1\% when varying $\mu$ between 1.25\,GeV and 5\,GeV.} We remark here
that for other values of the renormalization scale $\mu$ than $m_b$ it happens that the exact result can only
be approximated by the massless contribution when using a negative number of massless quark flavors. For
example, for $\mu=m_b/2$, the exact result can be approximated by setting $N_L=-1.32$ and $N_H=N_V=0$ in
(\ref{colorf}). Determining the effect of the new fermionic corrections on $\text{Br}(\bar B\to X_s\gamma)$ in
this case would require to repeat the interpolation procedure performed in \cite{Misiak:2004ew}. Since we expect 
that it will also be 1\% at most (when combining large-$\beta_0$ and massive quark loop corrections), that is
below the total uncertainty given in (\ref{estimate}), we postpone this to a forthcoming analysis which will
also include other contributions not considered so far.


\section{Summary}

In this paper we calculated the NNLO fermionic corrections to the total decay rate and the photon energy spectrum
induced by the interference of the electro- and chromomagnetic dipole operators. We confirmed the results for the
$O(\alpha_s^2\beta_0)$ terms given in \cite{Bieri:2003ue,Ligeti:1999ea} and also presented analytical results for
the contributions with massive bottom and charm quark loops. We expect that the combination of both the massless
and the massive quark loop contributions affects $\text{Br}(\bar B\to X_s\gamma)$ by 1\% at most.


\section*{\normalsize Acknowledgements}
\vspace*{-2mm}
I would like to thank Paolo Gambino, Christoph Greub and Mikolaj Misiak for comments on the final version of
this paper. I am also grateful to Mikolaj Misiak for detailed information about the numerical
effect of the $O(\alpha_s^2\beta_0)$ corrections in the $(O_7,O_8)$-interference on
$\text{Br}(\bar B\to X_s\gamma)$. This work was initiated when I was a postdoc at the Institute for Theoretical
Physics at the University of Bern and at that time it was supported in part by the EU Contract
No.~MRTN-CT-2006-035482, FLAVIAnet and by the Swiss National Foundation. Then, from October 2007 on, it was
supported in part by MIUR under contract 2004021808-009 and by a European Community's Marie-Curie Research
Training Network under contract MRTN-CT-2006-035505 `Tools and Precision Calculations for Physics Discoveries
at Colliders'. 


\appendix

\section{Two-particle cuts}\label{app_a}

In this appendix we specify the contributions of the 2-particle cuts entering the functions $Y^{(1)}$ and
$Y^{(2,i)}$ with $i=\mbox{\small NL}$, $\mbox{\small NH}$, $\mbox{\small NV}$. To this end we write
\begin{equation}
  Y^{(1)}(z_0,\mu) = Y^{(1)}_{2\text{-cuts}}(0,\mu) + Y^{(1)}_{3\text{-cuts}}(z_0,\mu)
\end{equation}
and
\begin{equation}
  Y^{(2)}(z_0,\mu) = Y^{(2)}_{2\text{-cuts}}(0,\mu) +
   Y^{(2)}_{3\text{-cuts}}(z_0,\mu) + Y^{(2)}_{4\text{-cuts}}(z_0,\mu)\,,
\end{equation}
with $Y^{(1)}_{n\text{-cuts}}$ and $Y^{(2)}_{n\text{-cuts}}$ summing all contributions of the $n$-particle cuts.
The contribution of the 2-particle cuts at $O(\alpha_s)$ reads
\begin{equation}
 Y^{(1)}_{2\text{-cuts}}(0,\mu) = \frac{2}{9}\left(33-2\pi^2\right) + \frac{16}{3}\,L_\mu\,,\label{Y2cuts}
\end{equation}
and those at $O(\alpha_s^2)$ are given by
\begin{align}
 Y^{(2,\mbox{{\tiny NL}})}_{2\text{-cuts}}(0,\mu) &= -\frac{16}{81}\left(157-8\pi^2\right) -
   \frac{16}{27}\left(47-2\pi^2\right)L_\mu-\frac{64}{9}\,L_\mu^2 +
   \frac{64}{9}\,\zeta_3\,,\label{Y2cutsnl}\\[3mm]
 Y^{(2,\mbox{{\tiny NH}})}_{2\text{-cuts}}(0,\mu) &= \frac{8}{81}\left(244-27\sqrt{3}\,\pi-61\pi^2\right) -
   \frac{16}{27}\left(47-2\pi ^2\right)L_\mu\nonumber\\[2mm]
 &\qquad -\frac{64}{9}\,L_\mu^2-\frac{64}{27}\,\zeta_3 +
   32\sqrt{3}\,\text{Cl}_2\left(\frac{\pi}{3}\right)\,,\label{Y2cutsnh}\\[3mm]
 Y^{(2,\mbox{{\tiny NV}})}_{2\text{-cuts}}(0,\mu) &=
   -\frac{16}{81}\left[157-279\rho-\pi^2\left(5+9\rho^2-42\rho^{3/2}\right)\right] -
   \frac{16}{27}\left(47-2\pi^2\right)L_\mu\nonumber\\[2mm]
 &\qquad -\frac{64}{9}\,L_\mu^2+\frac{8}{27}\left(19+20\rho\right)\ln\rho +
   \frac{16}{27}\left(8+9\rho^2\right)\ln^2\!\rho\nonumber\\[2mm]
 &\qquad -\frac{16}{27}\,\ln^3\!\rho-\frac{8}{9}\left(1-6\rho^2\right)\Phi_1(\rho) -
   \frac{8}{27}(19-46\rho)\,\Phi_2(\rho)\nonumber\\[2mm]
 &\qquad -\frac{32}{27}(13+14\rho)\,\Phi_3(\rho)-\frac{64}{9}\,\Phi_4(\rho) -
   \frac{32}{9}\,\ln\rho\,\text{Li}_2(1-\rho)\nonumber\\[2mm]
 &\qquad +\frac{32}{27}\left(5+9\rho^2+14\rho^{3/2}\right)\text{Li}_2(1-\rho) -
   \frac{1792}{27}\rho^{3/2}\,\text{Li}_2\left(1-\sqrt{\rho}\right)\nonumber\\[2mm]
 &\qquad +\frac{64}{9}\,\text{Li}_3(1-\rho)+\frac{64}{9}\,
   \text{Li}_3\left(1-\frac{1}{\rho}\right)\,.\label{Y2cutsnv}
\end{align}
In writing our results at $O(\alpha_s^2)$ we tacitly performed a splitting of $Y^{(2)}_{2\text{-cuts}}(0,\mu)$
into terms being proportional to certain combinations of colorfactors, in complete analogy to what we did in
(\ref{colorf}). We remark that all contributions given above are by themselves free of infrared and collinear
singularities, as well as independent of the gauge parameter entering the gluon propagator.


\section{Auxiliary functions}\label{app_b}

Here we collect the four auxiliary functions $\Phi_n(\rho)$ introduced in section \ref{sec_2}. They are defined as
follows,
\begin{align}
 \displaystyle\Phi_1(\rho) &= \theta(1-4\rho)\left[\ln^2y-\pi^2\right] -
   \theta(4\rho-1)\,\text{arccos}^2\left(1-\frac{1}{2\rho}\right)\,,\\[3mm]
 \displaystyle\Phi_2(\rho) &= \sqrt{|1-4\rho|}\left\{\theta(1-4\rho)\,\ln y -
   \theta(4\rho-1)\,\text{arccos}\left(1-\frac{1}{2\rho}\right)\right\}\,,\\[3mm]
 \displaystyle\Phi_3(\rho) &= \sqrt{|1-4\rho|}\,\bigg\{\theta(1-4\rho)\left[\text{Li}_2\left(-y\right) +
   \frac{1}{4}\ln^2y+\frac{\pi^2}{12}\right]\nonumber\\[2mm]
 &\hspace{3cm}
-\theta(4\rho-1)\,\text{Cl}_2\!\left(2\,\text{arcsin}
   \left(\frac{1}{2\sqrt{\rho}}\right)\right)\bigg\}\,,\\[3mm]
 \displaystyle\Phi_4(\rho) &= \theta(1-4\rho)\left[\text{Li}_3\left(-y\right) +
   \frac{1}{12}\ln^3y+\frac{\pi^2}{12}\ln y\right]\nonumber\\[2mm]
 &\qquad +\theta(4\rho-1)\,\text{Cl}_3\!\left(2\,\text{arcsin}\left(\frac{1}{2\sqrt{\rho}}\right)\right)\,,
\end{align}
where $\theta(z)$ is Heavyside's step function,
\begin{equation}
 y=\frac{1-\sqrt{\rule[0mm]{0mm}{3.1mm}1-4 \rho }}{1+\sqrt{\rule[0mm]{0mm}{3.1mm}1-4 \rho }}\,,
\end{equation}
and $\rho>0$. The definitions of the two Clausen functions appearing in the above given equations
read \cite{Lewin}
\begin{equation}
 \text{Cl}_2(z) = \text{Im}\!\left[\text{Li}_2\left(e^{i z}\right)\right]\,,
 \qquad\text{Cl}_3(z) = \text{Re}\!\left[\text{Li}_3\left(e^{-i z}\right)\right]\,,
\end{equation}
and those of the polylogarithms are given by
\begin{equation}
  \mbox{Li}_2(z) = -\int_0^z\!{\rm d}x\,\frac{\ln(1-x)}{x}\,,
  \qquad\mbox{Li}_3(z) = \int_0^z\!{\rm d}x\,\frac{\mbox{Li}_2(x)}{x}\,.
\end{equation}


\end{document}